\documentclass[twocolumn,superscriptaddress,amsmath,amssymb,aps,prl]{revtex4-1}

\usepackage{graphicx}
\usepackage{dcolumn}
\usepackage{bm}

\usepackage[colorlinks,urlcolor=blue,citecolor=blue,linkcolor=blue]{hyperref}

\begin{document}
	\title{Many-body Dynamics with Time-dependent Interaction}
	\affiliation{Key State Laboratory of Precision Spectroscopy, East China Normal University, Shanghai 200062, China}

	\author{Yanting Cheng}
	\affiliation{Institute for Advanced Study, Tsinghua University, Beijing 100084, China}

	\author{Zhe-Yu Shi}
	\email{zyshi@lps.ecnu.edu.cn}
	\affiliation{Key State Laboratory of Precision Spectroscopy, East China Normal University, Shanghai 200062, China}
	\date{\today}
	
	\begin{abstract}
	%Recent advances of cold atom experiment techniques such as the optical Feshbach resonance enable the experimental exploration of many-body dynamics of quantum systems with time-varying interaction strength.
	%Out-of-equilibrium dynamics of quantum systems...
	Recent advances in optical Feshbach resonance technique have enabled the experimental investigation of atomic gases with time-dependent interaction~\cite{clark2017collective,feng2019correlations}. In this work, we study the many-body dynamics of weakly interacting bosons subject with an arbitrary time varying scattering length.
	% Such system has recently been realized in trapped Bose-Einstein condensate via temporally modulated the interaction strength near an optical Feshbach resonance. 
	 By employing a variational ansatz% for the many-body wave function
	 , we derive an effective Hamiltonian that governs the dynamics of thermal particles. Crucially, we show that there exists a hidden symmetry in this Hamiltonian that can map the many-body dynamics to the precession of an SU(1,1) ``spin''% in an external ``magnetic field''
	 . As a demonstration, we calculate the situation where the scattering length is sinusoidally modulated. We show that the non-compactness of the SU(1,1) group naturally leads to solutions with exponentially growth of Bogoliubov modes and causes  instabilities.	
	\end{abstract}

	\maketitle
	
	The ability to accurately control various parameters in cold atomic gases allows the investigation of quantum matter under extreme conditions that are beyond reach in other physical systems.
	%The accurate controllability of parameters in cold atomic gases brings great potential for the investigation of quantum matter under extreme conditions that are beyond reach in other systems like neutron stars and condensed matter systems.
	Among these parameters, the tunable interaction strength is a key ingredient for many fascinating quantum phenomena such as the physics of BEC-BCS crossover~\cite{bourdel2004experimental,zwierlein2004condensation,chen2005bcs,giorgini2008theory}, superfluid to Mott insulator transition~\cite{greiner2002quantum,stoferle2004transition,folling2006formation,bakr2010probing} and the few-body Efimov effect~\cite{efimov1971weakly,efimov1970energy,kraemer2006evidence,braaten2006universality,naidon2017efimov}.
	
	One of the recent progress in controlling the interatomic interaction strength is the development of the optical Feshbach resonance technique~\cite{clark2017collective,feng2019correlations,theis2004tuning}. Comparing to the traditional magnetic Feshbach resonance which relies on tuning the magnetic field, the optical Feshbach resonance controls the interatomic interaction via changing the detuning and intensity of the optical field. Such difference allows the rapid and spatial modulations of the scattering length between atoms and thus enables the experimental investigation of a variety of exotic many-body dynamic systems. For example, the bose fireworks experiment recently carried out by the Chicago group shows that a bose condensate emits matter-wave jets and form striking fireworks patterns while subject to periodic modulated interactions~\cite{clark2017collective,fu2018density,wu2019dynamics}.
	
\begin{figure}[t]
		\centering
	\includegraphics[width=\linewidth]{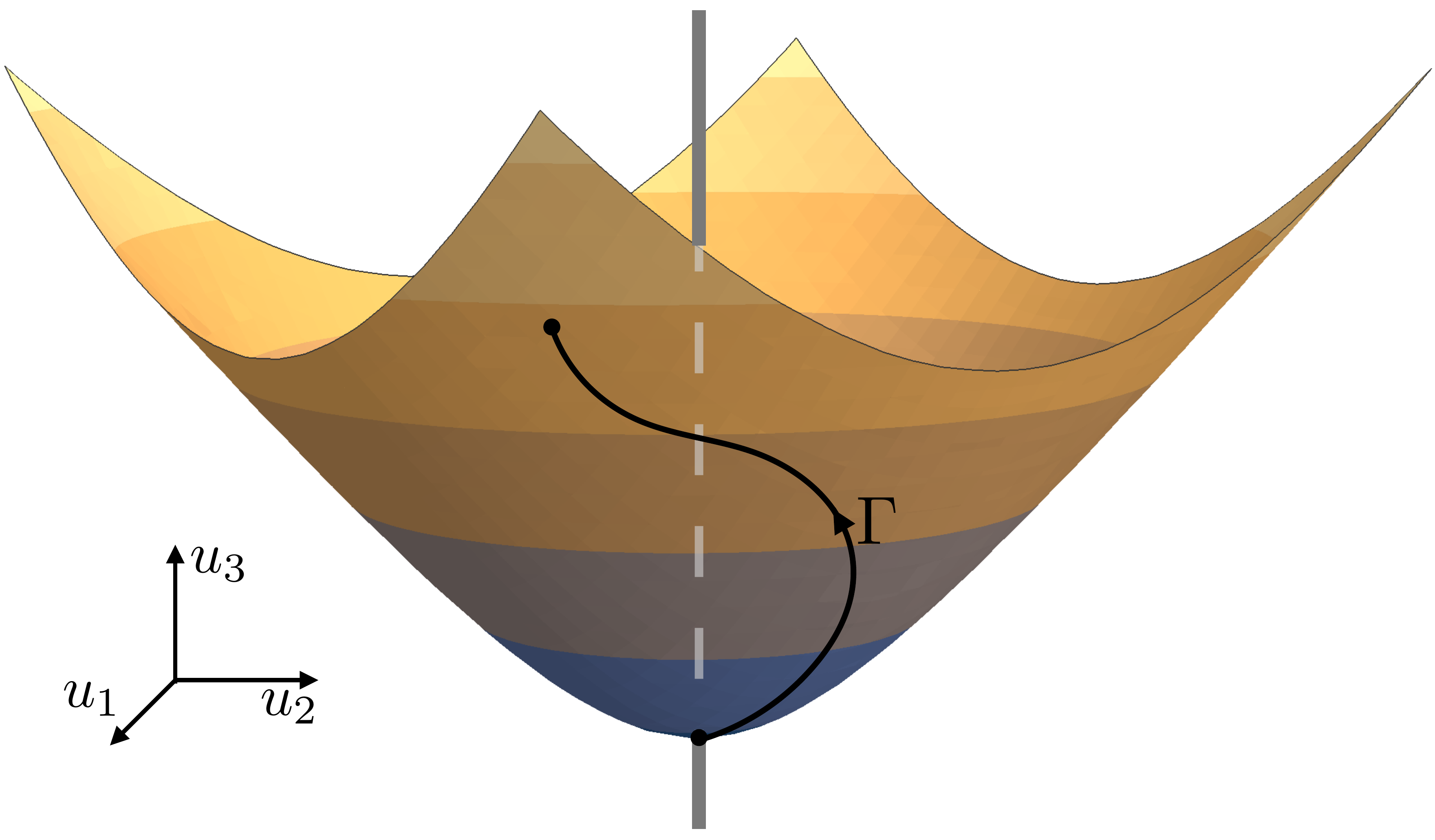}
	\caption{A schematic diagram showing the dynamics of the SU(1,1) spin on the ``Bloch'' hyperboloid. $\Gamma$ represents the trajectory of $\mathbf{u}(t)$. The Berry curvature on the hyperboloid is identical to the field of  a line of magnetic monopole represented by the gray line.}
	\label{fig1}	
	\end{figure}

	In this work, we focus on the dynamic problem of weakly interacting bose gas subject with an {\it arbitrary} time varying scattering length. In the low temperature limit, one might naively anticipate that the dynamics of the system could be described by a mean-field level Gross-Pitaevskii (GP) equation with time varying coupling constant $g(t)$. However, it is straightforward to show that the solution to the time-dependent GP equation is trivial as long as the system is initially in the ground state. This is closely related to the fact that the ground state solution ({\it i.e.} the saddle point) of a time-independent GP equation does not rely on the interaction strength $g$. Thus, it is necessary to go beyond the mean-field theory and consider the role of {\it quantum fluctuation}. In the corresponding static problem, the next order correction is known as the Lee-Huang-Yang correction which can be obtained by the Bogoliubov theory. Inspired by this correspondence, we propose a variational ansatz which accounts the next order quantum correction to the dynamic problem. We show that the dynamics of the variational wave function is governed by a Bogoliubov-like Hamiltonian. Crucially, we find that the Hamiltonian possesses a hidden SU(1,1) symmetry which not only allows an exact solution to the time-dependent Schr\"{o}dinger equation but also maps the dynamic problem to an SU(1,1) ``spin'' moving in a time-varying magnetic field. The SU(1,1) ``spin'' model closely resembles a normal SU(2) spin in an external field as its dynamics can be view as a point moving on a hyperboloid (see Fig.~\ref{fig1}) in parameter space which resembles an SU(2) Bloch sphere. To further demonstrate our method, we also calculate the dynamics of a system with periodically modulated scattering length. %We find that the non-compactness of the SU(1,1) group naturally leads to solutions with exponentially growing Bogoliubov modes which would eventually destroy the condensate. We also discuss the origin of the instability.
	  	%induced ones have the advantage of tuning the interaction strength 
	
	%In this work, we develop a beyond mean-field theory to describe the dynamics of weakly interacting bosons with time-varying interaction strength. By assuming the majority of the bosons condensed in the zero-momentum state, we find that the non-condensate part of the system can be well described by a Bogoliubov-like Hamilonian. Furthermore, we show that there is a hidden su(1,1) symmetry of the Hamiltonian that allows an exact solution to the dynamic problem.
	%In nowaday cold atom experiments, magnetic Feshbach resonance is the most common and effective method for tuning interaction strength between particles~\cite{inouye1998observation,courteille1998observation,chin2010feshbach}. Still, there is certain limitation that has restricted  the applications of this powerful technique. For example, because of the $s$-wave scattering length relies solely on the applied magnetic field, it is hard to modulate the system interaction spatially or temporally. On the other hand, since it is much easier to locally and rapidly modify optical fields, the optical Feshbach resonance technique was developed to overcome such limitations in magnetic Feshbach resonance.

	{\it Model.-- }We consider a Hamiltonian which describes a system of bosons interacting via short-range interaction,
	\begin{align}
 		H(t)=\sum_\mathbf{k}\epsilon_\mathbf{k}a_\mathbf{k}^\dagger a_\mathbf{k}+\frac{1}{2}g(t)\sum_{\mathbf{k,q,q'}}a_{\mathbf{q+k}}^\dagger a_{\mathbf{q'}}^\dagger a_{\mathbf{q}}a_{\mathbf{k'+q}}.
	\end{align}
	Here $a_\mathbf{k}^\dagger(a_\mathbf{k})$ are the bosonic creation (annihilation) operator with momentum $\mathbf{k}$ and mass $m$; $g(t)$ is an {\it arbitrary} time-varying coupling constant which is related to the $s$-wave scattering length $a_s$ by $g(t)={4\pi a_s(t)}/{m}$ (we set $\hbar$ and the volume of the system to $1$). The dynamic theory we develop in this work does {\it not} rely on the specific form of the dispersion $\epsilon_\mathbf{k}$ as long as the system has an inversion symmetry {\it i.e.} $\epsilon_\mathbf{k}=\epsilon_\mathbf{{-k}}$, and without loss of generality, we set $\epsilon_{\mathbf{0}}=0.$
	
	To proceed, we assume that the system is weakly interacting, such that during the dynamic process the majority of the bosons still condense in the zero-momentum state, {\it i.e.} $N_0(t)=\langle a^\dagger_\mathbf{0}a_\mathbf{0}\rangle\simeq N\gg1$. Therefore, one may approximate the time-dependent wave function by,
	\begin{align}
 		|\Psi(t)\rangle=|\psi(t)\rangle_{\mathbf{k\neq0}}\otimes e^{\sqrt{N_0}(a_\mathbf{0}^\dagger+a_\mathbf{0})}|0\rangle,\label{ansatz}
	\end{align}
	where the wave function $|\Psi(t)\rangle$ is decomposed into a product state of $|\psi(t)\rangle_\mathbf{k\neq0}$ which represents the state of non-condensed thermal bosons and a coherent state of $N_0$ condensed particles.
	
	To determine the ``best'' variational wave function $|\Psi(t)\rangle$, we use Frenknel's least action principle~\cite{frenkel1935wave,mclachlan1964variational} for dynamic systems and minimize the action $S=\int dt\langle\Psi(t)|i\partial_t-H(t)|\Psi(t)\rangle$~\cite{least_action}. This leads to a time-dependent Schr\"{o}dinger equation
	\begin{align}
 	i\partial_t|\psi(t)\rangle=H_B|\psi(t)\rangle,\label{t-d_sch}
	\end{align}
	with
	\begin{align}
 	H_B(t)&=\sum_\mathbf{k\neq0}(\epsilon_\mathbf{k}+g(t)n)a_\mathbf{k}^\dagger a_\mathbf{k}+\frac{g(t)n}{2}(a_\mathbf{k}^\dagger a_{\mathbf{-k}}^\dagger+h.c.)\nonumber\\&\qquad+\frac{g(t)nN}{2}+O(N^{-1/2}).
	\end{align}
	We see that the dynamics of thermal particles are governed by a Bogoliubov-like Hamiltonian $H_B(t)$.
	
	It is worth noting that simply diagonalizing $H_B(t)$ via Bogoliubov transformation does not solve the dynamic problem as its instantaneous eigen-state is not the solution to Eq.~\eqref{t-d_sch}. The solution to the dynamic problem actually relies on the hidden dynamic symmetry of Hamiltonian $H_B$.
	
	Note that the $\mathbf{k}$ part in $H_B$ only couples to $\mathbf{-k}$, and it can be rewritten as
	\begin{align}
 	H_B=\sum_{\mathbf{k}\neq0}\left[g(t)nA_1^\mathbf{k}+(\epsilon_\mathbf{k}+g(t)n)A_3^\mathbf{k}\right]+E_0.
	\end{align}
	Here $E_0=-\sum_\mathbf{k}(\epsilon_\mathbf{k}+gn)+{g(t)nN}/{2}$, $A^\mathbf{k}_1$ and $A^\mathbf{k}_3$ are defined as $A_1^\mathbf{k}=\frac{1}{2}(a_\mathbf{k}^\dagger a_{\mathbf{-k}}^\dagger+h.c.)$ and $A_3^\mathbf{k}=\frac{1}{2}(a_\mathbf{k}^\dagger a_\mathbf{k}+a_\mathbf{-k}a_\mathbf{-k}^\dagger)$.
	
	It was pointed out by Chen {\it et al.}~\cite{Chen2019} that $A_1^\mathbf{k}$ and $A_3^\mathbf{k}$ can fit into an su(1,1) algebra by including an extra operator $A_2^\mathbf{k}=\frac{1}{2i}(a_\mathbf{k}^\dagger a_{\mathbf{-k}}^\dagger-h.c.)$. Together with this operator, their commutators form a close algebra,
	\begin{align}
 	[A_1^\mathbf{k},A_2^\mathbf{k}]=-iA_3^\mathbf{k},\ [A_2^\mathbf{k},A_3^\mathbf{k}]=iA_1^\mathbf{k},\ [A_3^\mathbf{k},A_1^\mathbf{k}]=iA_2^\mathbf{k}.\label{com_rel}
	\end{align}
	Note that Eq.~\eqref{com_rel} differ from the common su(2) algebra of a spin system by a minus sign. As we will see in the following, there is a close resemblance between the dynamics of Bogoliubov systems and the dynamics of an SU(2) spin in a time-dependent magnetic field.

	{\it SU(1,1) spin model.- }Note that all the $(\mathbf{k,-k})$ subspaces with different $\mathbf{k}$ are decoupled, which allows us dealing with a pair of momenta at one time. For generality, in the following we will consider a model that consists of all the $A_i$ components,
	\begin{align}
 	H_h=\mathbf{h}\cdot\mathbf{A}^\mathbf{k}= h_1A_1^\mathbf{k}+h_2A_2^\mathbf{k}+h_3A_3^\mathbf{k}.
	\end{align}
	Here $\mathbf{h}=(h_1,h_2,h_3)^T$ is an arbitrary time-dependent vector and $H_h$ can be reduced to $H_B$ by letting $h_1=2g(t)n$, $h_2=0$ and $h_3=2(\epsilon_\mathbf{k}+g(t)n)$.
	
	The SU(1,1) symmetry leads to three time-dependent invariants for $H_h$. To see this, we consider operator $S$ in the form of $S=\sum_iu_i(t)A_i^\mathbf{k}$. In order to make $S$ a time-dependent invariant under $H_h$, we have
	\begin{align}
 	\frac{d}{dt}S=i[H_h,S]+\frac{\partial S}{\partial t}=0.
	\end{align}
	Thanks to the closed commutation relations, the above equation leads to a set of linear equations for $u_i$,
	\begin{align}
 	\dot{\mathbf{u}}=\left(\begin{array}{ccc}
0&-h_3&h_2\\
h_3&0&-h_1\\
h_2&-h_1&0
\end{array}
\right)\mathbf{u},\label{linear_eq}
	\end{align}
	with $\mathbf{u}=(u_1,u_2,u_3)^T$.
	
	Any $\mathbf{u}(t)$ satisfies Eq.~\eqref{linear_eq} represents an invariant $\mathbf{u\cdot A^k}$ for $H_h$. While there are three linear-independent solutions of this differential equation, which correspond to three independent invariants.
	
	%According to Lewis's theory for time-dependent system, the instantaneous eigen-state for $S(t)$ only diff
	%Once we have constructed the time-dependent invariants $S(t)$,

	{\it Remark on }$H_h$.-- %Here we would like to add a remark about the dynamics of su(1,1) Hamiltonain $H_h$.
	From Eq.~\eqref{linear_eq}, one can prove that $\Vert\mathbf{u}\Vert^2\equiv-u_1^2-u_2^2+u_3^2$ is a constant by showing that $\frac{d}{dt}\Vert \mathbf{u}\Vert=0$. This means that the three-dimensional vector $\mathbf{u}(t)$ is restricted on the surface of a hyperboloid defined by $-u_1^2-u_2^2+u_3^2=\text{const.}$. This may be viewed as the SU(1,1) analogue of the Bloch sphere in SU(2) spin case.
	
	Without loss of generality, we consider the solution of $\mathbf{u}(t)$ on the upper unit sheet of the hyperboloid as shown in Fig.~\ref{fig1}. The corresponding invariant can be parametrized as $S(t)=\mathbf{u}\cdot\mathbf{A}^\mathbf{k}=\sinh\theta\cos\phi A_1^\mathbf{k}+\sinh\theta\sin\phi A_2^\mathbf{k}+\cosh\theta A_3^\mathbf{k}$. Using the commutation relations in Eq.~\eqref{com_rel}, we can diagonalize it via the SU(1,1) rotation,
	\begin{align}
 	e^{iA_3^\mathbf{k}\phi}e^{iA_2^\mathbf{k}\theta} S(t)e^{-iA_2^\mathbf{k}\theta}e^{-iA_3^\mathbf{k}\phi}= A_3^\mathbf{k}.
	\end{align}
	Since $A_3^\mathbf{k}=\frac{1}{2}(n_\mathbf{k}+n_\mathbf{{-k}}+1)$, the eigenstates of $S(t)$ are thus parametrized by two integers $\mathbf{n}=(n_+,n_-)$ with $n_\pm$ the number of bosons in $\mathbf{k}$ and $\mathbf{-k}$ states. They are given by $|\mathbf{n}\rangle=|{n_+,n_-}\rangle=\frac{1}{\sqrt{n_+! n_-!}}e^{iA_3^\mathbf{k}\phi}e^{iA_2^\mathbf{k}\theta}a_\mathbf{k}^{\dagger n_+}a_\mathbf{-k}^{\dagger n_-}|0\rangle$.

	The instantaneous eigenstates of invariant $S(t)$ are useful because they are proportional to the solution to the Schr\"{o}dinger equation $|\Phi\rangle$. According to Lewis's theory for time-dependent invariants~\cite{lewis1969exact,lewis1967classical}, we have
	\begin{align}
 	|\Phi(t)\rangle=e^{-i\varphi(t)}|\mathbf{n}\rangle.\label{solution}
	\end{align}
	Here $|\Phi\rangle$ satisfies $[i\partial_t-H_h(t)]|\Phi(t)\rangle=0$. The phase $\varphi(t)$ contains a dynamical phase and a geometric phase with $\varphi(t)=\varphi_\text{dyn}(t)-\varphi_\text{g}(t)$,
	\begin{align}
 	\varphi_\text{dyn}=\int_{t_0}^td\tau\langle \mathbf{n}|H_h(\tau)|\mathbf{n}\rangle,\quad\varphi_\text{g}=i\int_{t_0}^td\tau\langle \mathbf{n}|\partial_\tau|\mathbf{n}\rangle.
	\end{align}
	
	Suppose the initial state of the system is the ground state of $\mathbf{h}_0\cdot\mathbf{A}^\mathbf{k}$, the initial condition for Eq.~\eqref{linear_eq} is then set as $\mathbf{u}(0)=\mathbf{h}_0$.  We can then obtain the solution of the time-dependent Schr\"{o}dinger equation by solving $\mathbf{u}(t)$ and substitute it into Eq.~\eqref{solution} with $n_+=n_-=0$.
		
	{\it Remarks on }$\varphi_\text{g}$.-- 
	%Before going to the detailed calculations, we first give a few remarks on the geometric phase $\varphi_\text{g}$.
	By changing variable $t$ to $\mathbf{u}$, we can show that the geometric phase $\varphi_g$ depends only on the trajectory of $\mathbf{u}$,
	\begin{align}
 	\varphi_\text{g}=i\int_\Gamma d\mathbf{u}\cdot\langle\mathbf{n}|\nabla_\mathbf{u}|\mathbf{n}\rangle=\int_\Gamma \mathcal{A}_\theta d\theta+\mathcal{A}_\phi d\phi,
	\end{align}
	where $\Gamma$ is the trajectory of $\mathbf{u}$ on the hyperboloid as shown in Fig.~\ref{fig1}. The Berry connection $\mathcal{A}_i$ is
	\begin{align}
 	\mathcal{A}_\theta&=i\langle\mathbf{n}|\partial_\theta|\mathbf{n}\rangle=0,\\
 	\mathcal{A}_\phi&=i\langle\mathbf{n}|\partial_\phi|\mathbf{n}\rangle=-C_\mathbf{n}\cosh\theta,
	\end{align}
	with charge $C_\mathbf{n}=({n_++n_-+1})/{2}$.
	
	As it is well known that the Berry curvature of an SU(2) spin is identical to the field of a Dirac monopole positioned at the center of the Bloch sphere. In the SU(1,1) dynamic theory, the Berry curvature in $\mathbf{u}$-space is given by $\nabla\times\mathbf{A}=C_\mathbf{n}\frac{\hat{e}_\rho}{\rho}$ with $\mathbf{\rho}=\sqrt{u_1^2+u_2^2}$ the radial coordinate and $\hat{e}_\rho=(u_1\hat{e}_1+u_2\hat{e}_2)/\rho$ is the unit vector along radial direction. This Berry curvature is equal to the field of a {\it line of Dirac monopoles} positioned on the $u_3$-axis with uniform linear density $d=C_\mathbf{n}/2$ as shown in Fig.~\ref{fig1}. The fact that the monopole line is infinitely long is a consequence of the non-compactness of the SU(1,1) group~\cite{monopole_line}.

	{\it  Bose gas with periodically driven $g(t)$.--} In the following, we consider a specific form of time-varying interaction strength with $g(t)=g_0+\delta g\sin\omega t$ and focus on the long-term behavior of the system. Such sinusoidal modulation is probably the most simple case and has already been implemented in several cold atom experiments~\cite{clark2017collective,feng2019correlations,nguyen2019parametric}.
	
	In the case of the weakly-interacting bose gas $h_1=2g(t)n$, $h_2=0$ and $h_3=2(\epsilon_\mathbf{k}+g(t)n)$, the  coupled linear equations~\eqref{linear_eq} can be further simplified into a single differential equation for $u_{31}\equiv u_3-u_1$~\cite{derivation_3rd_eq},
	\begin{align}
 	\dddot{u}_{31}+4\epsilon_\mathbf{k}(\epsilon_\mathbf{k}+2g(t)n)\dot{u}_{31}+4\epsilon_{\mathbf{k}}\dot{g}(t)nu_{31}=0.\label{3rd_eq}
	\end{align}
	One may check that the above equation is equivalent to the coupled equations~\eqref{linear_eq}.
	
	For the periodic driven case, the Floquet theorem asserts that the solution to Eq.~\eqref{linear_eq} must take the form of $\mathbf{u}(t)=e^{-iE_F t}\mathbf{p}(t)$ with $E_F$ the quasi-energy and $\mathbf{p}(t)$ a periodic function in $t$. 	
\begin{figure}[t]
	\centering
	\includegraphics[width=\linewidth]{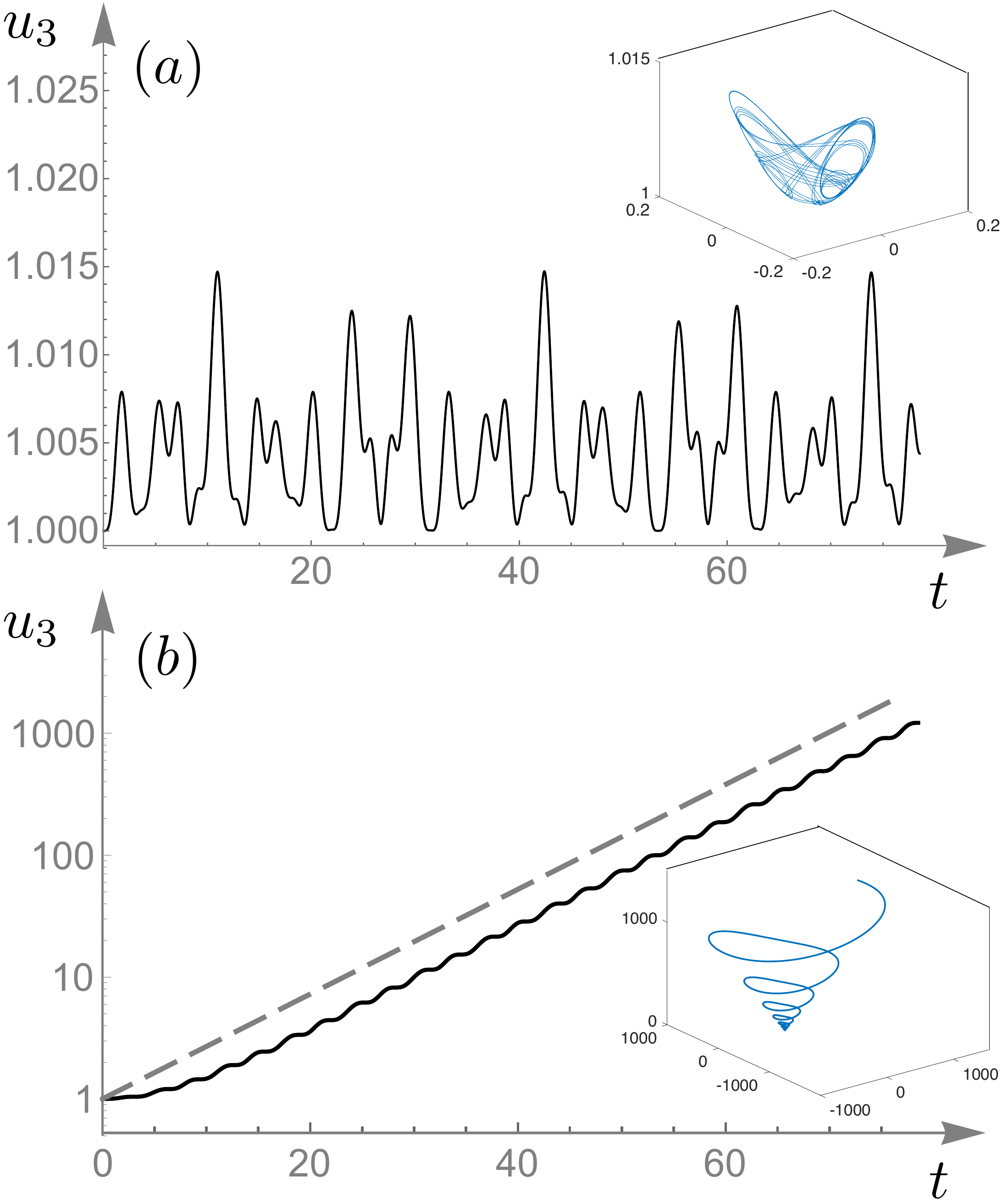}
	\caption{Sold lines: Typical long-term behavior for $u_3(t)$.  The time $t$ is plotted in unit of $1/\omega$. For both plots, we have $g_0=0$, $\delta g=0.1{\omega}/{n}$. In plot (a), we set $\epsilon_\mathbf{k}=1.2\omega$, which leads to a bounded oscillating behavior. In (b), we set $\epsilon_\mathbf{k}=0.5\omega$, and find that $u_3$ grows exponentially in the long term (The $y$-axes is in $log$-scale.). Dashed line: $e^{\lambda t}$ with $\lambda$ the Lyapunov exponent calculated by the Floquet theory. One can see its long-term trend nicely agrees with $u_3(t)$. The insets show actual trajectories of $\mathbf{u}(t)$ in both cases.}\label{fig2}
	\end{figure}
	
		\begin{figure*}[t]
	\includegraphics[width=1\textwidth]{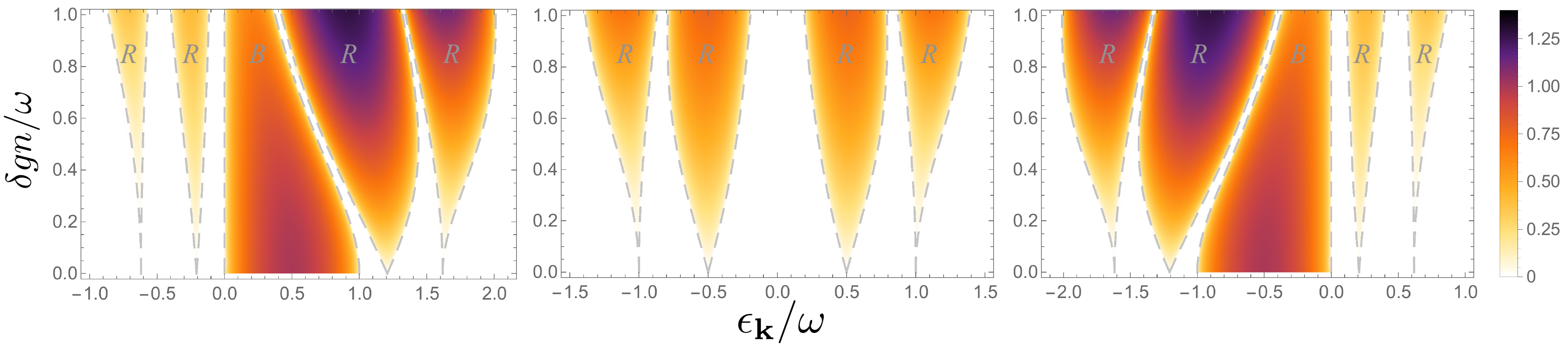}
	\caption{Stability diagram for Bose gas with oscillating interaction strength $g(t)=g_0+\delta g\sin\omega t$. From left to right: $g_0=-0.5\omega/n$, $g_0=0$, $g_0=0.5\omega/n$.
	The white area marks the stable region with vanishing Lyapunov exponent. Colored area marks the unstable regions. Dashed lines represent the transition curves that separate two regions. Lyapunov exponent is shown via the colormap. `B's and `R's in the instability lobes stand for Bogoliubov and resonance, which categorizes two different origins of the instabilities. }	\label{fig3}
	\end{figure*}
	The quasi-energy $E_F=\alpha+i\lambda$ is in general complex and its imaginary part controls the stability of the system. For a real quasi-energy, {\it i.e. $\lambda=0$}, the vector $\mathbf{u}$ is always bounded, meaning the condensate only emits a finite number of thermal particles with momentum $\pm\mathbf{k}$. On the other hand, if  the quasi-energy $E_F$ is complex, the $\mathbf{u}$ grows exponentially in the long term, meaning the condensate will keep emitting thermal particles until the variational wave function~\eqref{ansatz} breaks down. As one can see, the imaginary part $\lambda$ plays an important role of controlling the growth speed of the thermal modes, which can thus be interpreted as the Lyapunov exponent of the system.
	
	In Fig.~\ref{fig2}, we plot $u_3(t)$  for both cases by solving Eq.~\eqref{linear_eq} and show that $u_3$ indeed grows in the form of $e^{\lambda t}$. This is in contrast to the dynamics of an SU(2) spin as all the components of the SU(2) spin is bounded. As one can see from the insets, the exponentially growing solutions are related to the non-compactness of the SU(1,1) group, which is the main difference between the SU(1,1) and SU(2) groups.

	To calculate $\lambda$, we further show that the third order equation~\eqref{3rd_eq} is related to a second order one,
	\begin{align}
 	\ddot{v}+\epsilon_\mathbf{k}(\epsilon_\mathbf{k}+2g(t)n)v=0.\label{2nd_eq}
	\end{align}
	Namely, if $v_1$, $v_2$ are the two solutions of Eq.~\eqref{2nd_eq}, $u=v_1v_2$ is then the solution of Eq.~\eqref{3rd_eq}. Thus the three linear independent solutions for Eq.~\eqref{3rd_eq} are given by $v_1^2$, $v_1 v_2$ and $v_2^2$, with $v_1$, $v_2$ the linear independent solutions of Eq.~\eqref{2nd_eq}~\cite{deng2016observation,shi2017dynamic}.
	For $g(t)=g_0+\delta g\sin\omega t$, Eq.~\eqref{2nd_eq} reduces to a Mathieu's equation. The Mathieu's equation can be used to describe the classical dynamics of a parametric oscillator, whose long-term Lyapunov exponent~\cite{Lyapunov} may be calculated by the standard Whittaker-Hill's formula~\cite{mclachlan1951theory}.

	We plot the Lyapunov exponent $\lambda$ as a function of $\epsilon_\mathbf{k}$ and $\delta g$ in Fig.~\ref{fig3}. One can see that the system develops several instability lobes while turning on the modulation $\delta g$. These lobes are caused by two types of instability -- the resonance instability and the Bogoliubov instability. The resonance instability lobes emerge from $\sqrt{\epsilon_\mathbf{k}(\epsilon_\mathbf{k}+2g_0n)}=n\omega/2$ for small modulation strength $\delta g$ and keeps growing while increasing $\delta g$. Note that $\sqrt{\epsilon_\mathbf{k}(\epsilon_\mathbf{k}+2g_0n)}$ is the energy for Bogoliubov mode in the unperturbed system. This indicates that those instability appears because of the driven frequency is in resonance with two Bogoliubov excitations (one $\mathbf{k}$ and one $\mathbf{-k}$) of the unperturbed system. The Bogoliubov instability lobes exist even when there is no interaction strength modulation and shrink with increasing $\delta g$. They appear when $\epsilon_\mathbf{k}(\epsilon_\mathbf{k}+2g_0n)<0$, corresponding to the system has imaginary energy for Bogoliubov mode. Such instability is an intrinsic instability of the unperturbed system and hence be named Bogoliubov instability. The fact that the Bogoliubov instability lobes shrink with increasing $\delta g$ suggests that we may actually use the temporally modulated interaction to stablize condensates that are originally unstable with static interactions ({\it e.g.} bosons with attractive interaction).

%	Now we have constructed the time-dependent invariants $S(t)$, 

	To conclude, we have developed a beyond mean-field theory to describe the dynamics of weakly interacting bosons with time-varying interaction strength. By assuming the majority of the bosons is condensed in the ground state, we found that the non-condensate part of the system can be well described by a Bogoliubov-like Hamilonian. Furthermore, by identifying a hidden SU(1,1) symmetry of the system, we show that the dynamic problem of bosons can be mapped to the problem of an SU(1,1) spin in a time-varying magnetic field.  We explicitly constructed the time-dependent invariants of this system which gives the exact solution to the original time-dependent Schr\"{o}dinger equation. Interestingly, the Berry curvature of the SU(1,1) spin is found to be identical to the field of a line of Dirac monopoles. Experiments that can generate such gauge field in a BEC has been proposed for years but not yet realized~\cite{conduit2012line}. Thus the model we described in this work might provide an alternative and feasiable method to create and simulate such a novel configuration of gauge fields. %As an example, we calculated the dynamics of weakly interacting bosons with sinusoidally modulated interaction strength and discussed its stability properties via further mapping the problem to a classic parametric oscillator.
	
	We acknowledge fruitful discussions with Hui Zhai, Wei Zheng, Zhigang Wu, Meera Parish and Jesper Levinsen.

	%{\it Note added.-} ``{\it `Forty-two'}, said Deep Thought, with infinite majesty and calm.''~\cite{adams2017hitchhiker}

	\bibliography{references}

\begin{thebibliography}{32}%
\makeatletter
\providecommand \@ifxundefined [1]{%
 \@ifx{#1\undefined}
}%
\providecommand \@ifnum [1]{%
 \ifnum #1\expandafter \@firstoftwo
 \else \expandafter \@secondoftwo
 \fi
}%
\providecommand \@ifx [1]{%
 \ifx #1\expandafter \@firstoftwo
 \else \expandafter \@secondoftwo
 \fi
}%
\providecommand \natexlab [1]{#1}%
\providecommand \enquote  [1]{``#1''}%
\providecommand \bibnamefont  [1]{#1}%
\providecommand \bibfnamefont [1]{#1}%
\providecommand \citenamefont [1]{#1}%
\providecommand \href@noop [0]{\@secondoftwo}%
\providecommand \href [0]{\begingroup \@sanitize@url \@href}%
\providecommand \@href[1]{\@@startlink{#1}\@@href}%
\providecommand \@@href[1]{\endgroup#1\@@endlink}%
\providecommand \@sanitize@url [0]{\catcode `\\12\catcode `\$12\catcode
  `\&12\catcode `\#12\catcode `\^12\catcode `\_12\catcode `\%12\relax}%
\providecommand \@@startlink[1]{}%
\providecommand \@@endlink[0]{}%
\providecommand \url  [0]{\begingroup\@sanitize@url \@url }%
\providecommand \@url [1]{\endgroup\@href {#1}{\urlprefix }}%
\providecommand \urlprefix  [0]{URL }%
\providecommand \Eprint [0]{\href }%
\providecommand \doibase [0]{http://dx.doi.org/}%
\providecommand \selectlanguage [0]{\@gobble}%
\providecommand \bibinfo  [0]{\@secondoftwo}%
\providecommand \bibfield  [0]{\@secondoftwo}%
\providecommand \translation [1]{[#1]}%
\providecommand \BibitemOpen [0]{}%
\providecommand \bibitemStop [0]{}%
\providecommand \bibitemNoStop [0]{.\EOS\space}%
\providecommand \EOS [0]{\spacefactor3000\relax}%
\providecommand \BibitemShut  [1]{\csname bibitem#1\endcsname}%
\let\auto@bib@innerbib\@empty
%</preamble>
\bibitem [{\citenamefont {Clark}\ \emph {et~al.}(2017)\citenamefont {Clark},
  \citenamefont {Gaj}, \citenamefont {Feng},\ and\ \citenamefont
  {Chin}}]{clark2017collective}%
  \BibitemOpen
  \bibfield  {author} {\bibinfo {author} {\bibfnamefont {L.~W.}\ \bibnamefont
  {Clark}}, \bibinfo {author} {\bibfnamefont {A.}~\bibnamefont {Gaj}}, \bibinfo
  {author} {\bibfnamefont {L.}~\bibnamefont {Feng}}, \ and\ \bibinfo {author}
  {\bibfnamefont {C.}~\bibnamefont {Chin}},\ }\bibfield  {title} {\bibinfo
  {title} {\emph {Collective emission of matter-wave jets from driven
  Bose--Einstein condensates}},\ }\href@noop {} {\bibfield  {journal} {\bibinfo
   {journal} {Nature}\ }\textbf {\bibinfo {volume} {551}},\ \bibinfo {pages}
  {356} (\bibinfo {year} {2017})}\BibitemShut {NoStop}%
\bibitem [{\citenamefont {Feng}\ \emph {et~al.}(2019)\citenamefont {Feng},
  \citenamefont {Hu}, \citenamefont {Clark},\ and\ \citenamefont
  {Chin}}]{feng2019correlations}%
  \BibitemOpen
  \bibfield  {author} {\bibinfo {author} {\bibfnamefont {L.}~\bibnamefont
  {Feng}}, \bibinfo {author} {\bibfnamefont {J.}~\bibnamefont {Hu}}, \bibinfo
  {author} {\bibfnamefont {L.~W.}\ \bibnamefont {Clark}}, \ and\ \bibinfo
  {author} {\bibfnamefont {C.}~\bibnamefont {Chin}},\ }\bibfield  {title}
  {\bibinfo {title} {\emph {Correlations in high-harmonic generation of
  matter-wave jets revealed by pattern recognition}},\ }\href@noop {}
  {\bibfield  {journal} {\bibinfo  {journal} {Science}\ }\textbf {\bibinfo
  {volume} {363}},\ \bibinfo {pages} {521} (\bibinfo {year}
  {2019})}\BibitemShut {NoStop}%
\bibitem [{\citenamefont {Bourdel}\ \emph {et~al.}(2004)\citenamefont
  {Bourdel}, \citenamefont {Khaykovich}, \citenamefont {Cubizolles},
  \citenamefont {Zhang}, \citenamefont {Chevy}, \citenamefont {Teichmann},
  \citenamefont {Tarruell}, \citenamefont {Kokkelmans},\ and\ \citenamefont
  {Salomon}}]{bourdel2004experimental}%
  \BibitemOpen
  \bibfield  {author} {\bibinfo {author} {\bibfnamefont {T.}~\bibnamefont
  {Bourdel}}, \bibinfo {author} {\bibfnamefont {L.}~\bibnamefont {Khaykovich}},
  \bibinfo {author} {\bibfnamefont {J.}~\bibnamefont {Cubizolles}}, \bibinfo
  {author} {\bibfnamefont {J.}~\bibnamefont {Zhang}}, \bibinfo {author}
  {\bibfnamefont {F.}~\bibnamefont {Chevy}}, \bibinfo {author} {\bibfnamefont
  {M.}~\bibnamefont {Teichmann}}, \bibinfo {author} {\bibfnamefont
  {L.}~\bibnamefont {Tarruell}}, \bibinfo {author} {\bibfnamefont
  {S.}~\bibnamefont {Kokkelmans}}, \ and\ \bibinfo {author} {\bibfnamefont
  {C.}~\bibnamefont {Salomon}},\ }\bibfield  {title} {\bibinfo {title} {\emph
  {Experimental study of the BEC-BCS crossover region in lithium 6}},\
  }\href@noop {} {\bibfield  {journal} {\bibinfo  {journal} {Physical Review
  Letters}\ }\textbf {\bibinfo {volume} {93}},\ \bibinfo {pages} {050401}
  (\bibinfo {year} {2004})}\BibitemShut {NoStop}%
\bibitem [{\citenamefont {Zwierlein}\ \emph {et~al.}(2004)\citenamefont
  {Zwierlein}, \citenamefont {Stan}, \citenamefont {Schunck}, \citenamefont
  {Raupach}, \citenamefont {Kerman},\ and\ \citenamefont
  {Ketterle}}]{zwierlein2004condensation}%
  \BibitemOpen
  \bibfield  {author} {\bibinfo {author} {\bibfnamefont {M.}~\bibnamefont
  {Zwierlein}}, \bibinfo {author} {\bibfnamefont {C.}~\bibnamefont {Stan}},
  \bibinfo {author} {\bibfnamefont {C.}~\bibnamefont {Schunck}}, \bibinfo
  {author} {\bibfnamefont {S.}~\bibnamefont {Raupach}}, \bibinfo {author}
  {\bibfnamefont {A.}~\bibnamefont {Kerman}}, \ and\ \bibinfo {author}
  {\bibfnamefont {W.}~\bibnamefont {Ketterle}},\ }\bibfield  {title} {\bibinfo
  {title} {\emph {Condensation of pairs of fermionic atoms near a Feshbach
  resonance}},\ }\href@noop {} {\bibfield  {journal} {\bibinfo  {journal}
  {Physical Review Letters}\ }\textbf {\bibinfo {volume} {92}},\ \bibinfo
  {pages} {120403} (\bibinfo {year} {2004})}\BibitemShut {NoStop}%
\bibitem [{\citenamefont {Chen}\ \emph {et~al.}(2005)\citenamefont {Chen},
  \citenamefont {Stajic}, \citenamefont {Tan},\ and\ \citenamefont
  {Levin}}]{chen2005bcs}%
  \BibitemOpen
  \bibfield  {author} {\bibinfo {author} {\bibfnamefont {Q.}~\bibnamefont
  {Chen}}, \bibinfo {author} {\bibfnamefont {J.}~\bibnamefont {Stajic}},
  \bibinfo {author} {\bibfnamefont {S.}~\bibnamefont {Tan}}, \ and\ \bibinfo
  {author} {\bibfnamefont {K.}~\bibnamefont {Levin}},\ }\bibfield  {title}
  {\bibinfo {title} {\emph {BCS--BEC crossover: From high temperature
  superconductors to ultracold superfluids}},\ }\href@noop {} {\bibfield
  {journal} {\bibinfo  {journal} {Physics Reports}\ }\textbf {\bibinfo {volume}
  {412}},\ \bibinfo {pages} {1} (\bibinfo {year} {2005})}\BibitemShut {NoStop}%
\bibitem [{\citenamefont {Giorgini}\ \emph {et~al.}(2008)\citenamefont
  {Giorgini}, \citenamefont {Pitaevskii},\ and\ \citenamefont
  {Stringari}}]{giorgini2008theory}%
  \BibitemOpen
  \bibfield  {author} {\bibinfo {author} {\bibfnamefont {S.}~\bibnamefont
  {Giorgini}}, \bibinfo {author} {\bibfnamefont {L.~P.}\ \bibnamefont
  {Pitaevskii}}, \ and\ \bibinfo {author} {\bibfnamefont {S.}~\bibnamefont
  {Stringari}},\ }\bibfield  {title} {\bibinfo {title} {\emph {Theory of
  ultracold atomic Fermi gases}},\ }\href@noop {} {\bibfield  {journal}
  {\bibinfo  {journal} {Reviews of Modern Physics}\ }\textbf {\bibinfo {volume}
  {80}},\ \bibinfo {pages} {1215} (\bibinfo {year} {2008})}\BibitemShut
  {NoStop}%
\bibitem [{\citenamefont {Greiner}\ \emph {et~al.}(2002)\citenamefont
  {Greiner}, \citenamefont {Mandel}, \citenamefont {Esslinger}, \citenamefont
  {H{\"a}nsch},\ and\ \citenamefont {Bloch}}]{greiner2002quantum}%
  \BibitemOpen
  \bibfield  {author} {\bibinfo {author} {\bibfnamefont {M.}~\bibnamefont
  {Greiner}}, \bibinfo {author} {\bibfnamefont {O.}~\bibnamefont {Mandel}},
  \bibinfo {author} {\bibfnamefont {T.}~\bibnamefont {Esslinger}}, \bibinfo
  {author} {\bibfnamefont {T.~W.}\ \bibnamefont {H{\"a}nsch}}, \ and\ \bibinfo
  {author} {\bibfnamefont {I.}~\bibnamefont {Bloch}},\ }\bibfield  {title}
  {\bibinfo {title} {\emph {Quantum phase transition from a superfluid to a
  Mott insulator in a gas of ultracold atoms}},\ }\href@noop {} {\bibfield
  {journal} {\bibinfo  {journal} {nature}\ }\textbf {\bibinfo {volume} {415}},\
  \bibinfo {pages} {39} (\bibinfo {year} {2002})}\BibitemShut {NoStop}%
\bibitem [{\citenamefont {St{\"o}ferle}\ \emph {et~al.}(2004)\citenamefont
  {St{\"o}ferle}, \citenamefont {Moritz}, \citenamefont {Schori}, \citenamefont
  {K{\"o}hl},\ and\ \citenamefont {Esslinger}}]{stoferle2004transition}%
  \BibitemOpen
  \bibfield  {author} {\bibinfo {author} {\bibfnamefont {T.}~\bibnamefont
  {St{\"o}ferle}}, \bibinfo {author} {\bibfnamefont {H.}~\bibnamefont
  {Moritz}}, \bibinfo {author} {\bibfnamefont {C.}~\bibnamefont {Schori}},
  \bibinfo {author} {\bibfnamefont {M.}~\bibnamefont {K{\"o}hl}}, \ and\
  \bibinfo {author} {\bibfnamefont {T.}~\bibnamefont {Esslinger}},\ }\bibfield
  {title} {\bibinfo {title} {\emph {Transition from a strongly interacting 1D
  superfluid to a Mott insulator}},\ }\href@noop {} {\bibfield  {journal}
  {\bibinfo  {journal} {Physical review letters}\ }\textbf {\bibinfo {volume}
  {92}},\ \bibinfo {pages} {130403} (\bibinfo {year} {2004})}\BibitemShut
  {NoStop}%
\bibitem [{\citenamefont {F{\"o}lling}\ \emph {et~al.}(2006)\citenamefont
  {F{\"o}lling}, \citenamefont {Widera}, \citenamefont {M{\"u}ller},
  \citenamefont {Gerbier},\ and\ \citenamefont {Bloch}}]{folling2006formation}%
  \BibitemOpen
  \bibfield  {author} {\bibinfo {author} {\bibfnamefont {S.}~\bibnamefont
  {F{\"o}lling}}, \bibinfo {author} {\bibfnamefont {A.}~\bibnamefont {Widera}},
  \bibinfo {author} {\bibfnamefont {T.}~\bibnamefont {M{\"u}ller}}, \bibinfo
  {author} {\bibfnamefont {F.}~\bibnamefont {Gerbier}}, \ and\ \bibinfo
  {author} {\bibfnamefont {I.}~\bibnamefont {Bloch}},\ }\bibfield  {title}
  {\bibinfo {title} {\emph {Formation of spatial shell structure in the
  superfluid to Mott insulator transition}},\ }\href@noop {} {\bibfield
  {journal} {\bibinfo  {journal} {Physical Review Letters}\ }\textbf {\bibinfo
  {volume} {97}},\ \bibinfo {pages} {060403} (\bibinfo {year}
  {2006})}\BibitemShut {NoStop}%
\bibitem [{\citenamefont {Bakr}\ \emph {et~al.}(2010)\citenamefont {Bakr},
  \citenamefont {Peng}, \citenamefont {Tai}, \citenamefont {Ma}, \citenamefont
  {Simon}, \citenamefont {Gillen}, \citenamefont {Foelling}, \citenamefont
  {Pollet},\ and\ \citenamefont {Greiner}}]{bakr2010probing}%
  \BibitemOpen
  \bibfield  {author} {\bibinfo {author} {\bibfnamefont {W.~S.}\ \bibnamefont
  {Bakr}}, \bibinfo {author} {\bibfnamefont {A.}~\bibnamefont {Peng}}, \bibinfo
  {author} {\bibfnamefont {M.~E.}\ \bibnamefont {Tai}}, \bibinfo {author}
  {\bibfnamefont {R.}~\bibnamefont {Ma}}, \bibinfo {author} {\bibfnamefont
  {J.}~\bibnamefont {Simon}}, \bibinfo {author} {\bibfnamefont {J.~I.}\
  \bibnamefont {Gillen}}, \bibinfo {author} {\bibfnamefont {S.}~\bibnamefont
  {Foelling}}, \bibinfo {author} {\bibfnamefont {L.}~\bibnamefont {Pollet}}, \
  and\ \bibinfo {author} {\bibfnamefont {M.}~\bibnamefont {Greiner}},\
  }\bibfield  {title} {\bibinfo {title} {\emph {Probing the
  superfluid--to--Mott insulator transition at the single-atom level}},\
  }\href@noop {} {\bibfield  {journal} {\bibinfo  {journal} {Science}\ }\textbf
  {\bibinfo {volume} {329}},\ \bibinfo {pages} {547} (\bibinfo {year}
  {2010})}\BibitemShut {NoStop}%
\bibitem [{\citenamefont {Efimov}(1971)}]{efimov1971weakly}%
  \BibitemOpen
  \bibfield  {author} {\bibinfo {author} {\bibfnamefont {V.}~\bibnamefont
  {Efimov}},\ }\bibfield  {title} {\bibinfo {title} {\emph {Weakly-bound states
  of three resonantly-interacting particles}},\ }\href@noop {} {\bibfield
  {journal} {\bibinfo  {journal} {Sov. J. Nucl. Phys.}\ }\textbf {\bibinfo
  {volume} {12}},\ \bibinfo {pages} {101} (\bibinfo {year} {1971})}\BibitemShut
  {NoStop}%
\bibitem [{\citenamefont {Efimov}(1970)}]{efimov1970energy}%
  \BibitemOpen
  \bibfield  {author} {\bibinfo {author} {\bibfnamefont {V.}~\bibnamefont
  {Efimov}},\ }\bibfield  {title} {\bibinfo {title} {\emph {Energy levels
  arising from resonant two-body forces in a three-body system}},\ }\href@noop
  {} {\bibfield  {journal} {\bibinfo  {journal} {Phys. Lett. B}\ }\textbf
  {\bibinfo {volume} {33}},\ \bibinfo {pages} {563} (\bibinfo {year}
  {1970})}\BibitemShut {NoStop}%
\bibitem [{\citenamefont {Kraemer}\ \emph {et~al.}(2006)\citenamefont
  {Kraemer}, \citenamefont {Mark}, \citenamefont {Waldburger}, \citenamefont
  {Danzl}, \citenamefont {Chin}, \citenamefont {Engeser}, \citenamefont
  {Lange}, \citenamefont {Pilch}, \citenamefont {Jaakkola}, \citenamefont
  {N{\"a}gerl} \emph {et~al.}}]{kraemer2006evidence}%
  \BibitemOpen
  \bibfield  {author} {\bibinfo {author} {\bibfnamefont {T.}~\bibnamefont
  {Kraemer}}, \bibinfo {author} {\bibfnamefont {M.}~\bibnamefont {Mark}},
  \bibinfo {author} {\bibfnamefont {P.}~\bibnamefont {Waldburger}}, \bibinfo
  {author} {\bibfnamefont {J.~G.}\ \bibnamefont {Danzl}}, \bibinfo {author}
  {\bibfnamefont {C.}~\bibnamefont {Chin}}, \bibinfo {author} {\bibfnamefont
  {B.}~\bibnamefont {Engeser}}, \bibinfo {author} {\bibfnamefont {A.~D.}\
  \bibnamefont {Lange}}, \bibinfo {author} {\bibfnamefont {K.}~\bibnamefont
  {Pilch}}, \bibinfo {author} {\bibfnamefont {A.}~\bibnamefont {Jaakkola}},
  \bibinfo {author} {\bibfnamefont {H.-C.}\ \bibnamefont {N{\"a}gerl}},  \emph
  {et~al.},\ }\bibfield  {title} {\bibinfo {title} {\emph {Evidence for Efimov
  quantum states in an ultracold gas of caesium atoms}},\ }\href@noop {}
  {\bibfield  {journal} {\bibinfo  {journal} {Nature}\ }\textbf {\bibinfo
  {volume} {440}},\ \bibinfo {pages} {315} (\bibinfo {year}
  {2006})}\BibitemShut {NoStop}%
\bibitem [{\citenamefont {Braaten}\ and\ \citenamefont
  {Hammer}(2006)}]{braaten2006universality}%
  \BibitemOpen
  \bibfield  {author} {\bibinfo {author} {\bibfnamefont {E.}~\bibnamefont
  {Braaten}}\ and\ \bibinfo {author} {\bibfnamefont {H.-W.}\ \bibnamefont
  {Hammer}},\ }\bibfield  {title} {\bibinfo {title} {\emph {Universality in
  few-body systems with large scattering length}},\ }\href@noop {} {\bibfield
  {journal} {\bibinfo  {journal} {Phys. Rep.}\ }\textbf {\bibinfo {volume}
  {428}},\ \bibinfo {pages} {259} (\bibinfo {year} {2006})}\BibitemShut
  {NoStop}%
\bibitem [{\citenamefont {Naidon}\ and\ \citenamefont
  {Endo}(2017)}]{naidon2017efimov}%
  \BibitemOpen
  \bibfield  {author} {\bibinfo {author} {\bibfnamefont {P.}~\bibnamefont
  {Naidon}}\ and\ \bibinfo {author} {\bibfnamefont {S.}~\bibnamefont {Endo}},\
  }\bibfield  {title} {\bibinfo {title} {\emph {Efimov physics: a review}},\
  }\href@noop {} {\bibfield  {journal} {\bibinfo  {journal} {Rep. Prog. Phys.}\
  }\textbf {\bibinfo {volume} {80}},\ \bibinfo {pages} {056001} (\bibinfo
  {year} {2017})}\BibitemShut {NoStop}%
\bibitem [{\citenamefont {Theis}\ \emph {et~al.}(2004)\citenamefont {Theis},
  \citenamefont {Thalhammer}, \citenamefont {Winkler}, \citenamefont {Hellwig},
  \citenamefont {Ruff}, \citenamefont {Grimm},\ and\ \citenamefont
  {Denschlag}}]{theis2004tuning}%
  \BibitemOpen
  \bibfield  {author} {\bibinfo {author} {\bibfnamefont {M.}~\bibnamefont
  {Theis}}, \bibinfo {author} {\bibfnamefont {G.}~\bibnamefont {Thalhammer}},
  \bibinfo {author} {\bibfnamefont {K.}~\bibnamefont {Winkler}}, \bibinfo
  {author} {\bibfnamefont {M.}~\bibnamefont {Hellwig}}, \bibinfo {author}
  {\bibfnamefont {G.}~\bibnamefont {Ruff}}, \bibinfo {author} {\bibfnamefont
  {R.}~\bibnamefont {Grimm}}, \ and\ \bibinfo {author} {\bibfnamefont {J.~H.}\
  \bibnamefont {Denschlag}},\ }\bibfield  {title} {\bibinfo {title} {\emph
  {Tuning the scattering length with an optically induced Feshbach
  resonance}},\ }\href@noop {} {\bibfield  {journal} {\bibinfo  {journal}
  {Physical Review Letters}\ }\textbf {\bibinfo {volume} {93}},\ \bibinfo
  {pages} {123001} (\bibinfo {year} {2004})}\BibitemShut {NoStop}%
\bibitem [{\citenamefont {Fu}\ \emph {et~al.}(2018)\citenamefont {Fu},
  \citenamefont {Feng}, \citenamefont {Anderson}, \citenamefont {Clark},
  \citenamefont {Hu}, \citenamefont {Andrade}, \citenamefont {Chin},\ and\
  \citenamefont {Levin}}]{fu2018density}%
  \BibitemOpen
  \bibfield  {author} {\bibinfo {author} {\bibfnamefont {H.}~\bibnamefont
  {Fu}}, \bibinfo {author} {\bibfnamefont {L.}~\bibnamefont {Feng}}, \bibinfo
  {author} {\bibfnamefont {B.~M.}\ \bibnamefont {Anderson}}, \bibinfo {author}
  {\bibfnamefont {L.~W.}\ \bibnamefont {Clark}}, \bibinfo {author}
  {\bibfnamefont {J.}~\bibnamefont {Hu}}, \bibinfo {author} {\bibfnamefont
  {J.~W.}\ \bibnamefont {Andrade}}, \bibinfo {author} {\bibfnamefont
  {C.}~\bibnamefont {Chin}}, \ and\ \bibinfo {author} {\bibfnamefont
  {K.}~\bibnamefont {Levin}},\ }\bibfield  {title} {\bibinfo {title} {\emph
  {Density waves and jet emission asymmetry in Bose Fireworks}},\ }\href@noop
  {} {\bibfield  {journal} {\bibinfo  {journal} {Physical review letters}\
  }\textbf {\bibinfo {volume} {121}},\ \bibinfo {pages} {243001} (\bibinfo
  {year} {2018})}\BibitemShut {NoStop}%
\bibitem [{\citenamefont {Wu}\ and\ \citenamefont
  {Zhai}(2019)}]{wu2019dynamics}%
  \BibitemOpen
  \bibfield  {author} {\bibinfo {author} {\bibfnamefont {Z.}~\bibnamefont
  {Wu}}\ and\ \bibinfo {author} {\bibfnamefont {H.}~\bibnamefont {Zhai}},\
  }\bibfield  {title} {\bibinfo {title} {\emph {Dynamics and density
  correlations in matter-wave jet emission of a driven condensate}},\
  }\href@noop {} {\bibfield  {journal} {\bibinfo  {journal} {Physical Review
  A}\ }\textbf {\bibinfo {volume} {99}},\ \bibinfo {pages} {063624} (\bibinfo
  {year} {2019})}\BibitemShut {NoStop}%
\bibitem [{\citenamefont {Frenkel}(1935)}]{frenkel1935wave}%
  \BibitemOpen
  \bibfield  {author} {\bibinfo {author} {\bibfnamefont {{\^A}.~I.}\
  \bibnamefont {Frenkel}},\ }\bibfield  {title} {\bibinfo {title} {\emph {Wave
  Mechanics; Advanced General Theory}},\ }\href@noop {} {\bibfield  {journal}
  {\bibinfo  {journal} {Bull. Amer. Math. Soc}\ }\textbf {\bibinfo {volume}
  {41}},\ \bibinfo {pages} {776} (\bibinfo {year} {1935})}\BibitemShut
  {NoStop}%
\bibitem [{\citenamefont {McLachlan}(1964)}]{mclachlan1964variational}%
  \BibitemOpen
  \bibfield  {author} {\bibinfo {author} {\bibfnamefont {A.}~\bibnamefont
  {McLachlan}},\ }\bibfield  {title} {\bibinfo {title} {\emph {A variational
  solution of the time-dependent Schrodinger equation}},\ }\href@noop {}
  {\bibfield  {journal} {\bibinfo  {journal} {Molecular Physics}\ }\textbf
  {\bibinfo {volume} {8}},\ \bibinfo {pages} {39} (\bibinfo {year}
  {1964})}\BibitemShut {NoStop}%
\bibitem [{lea()}]{least_action}%
  \BibitemOpen
  \href@noop {} {}\bibinfo {note} {Note that the least action principle leads
  to the exact time-dependent Schr\"{o}dinger equation if we put no restriction
  on the wave function $|\Psi(t)\rangle$.}\BibitemShut {Stop}%
\bibitem [{\citenamefont {Chen}\ \emph {et~al.}()\citenamefont {Chen},
  \citenamefont {Zhang}, \citenamefont {Zheng}, \citenamefont {Wu},\ and\
  \citenamefont {Zhai}}]{Chen2019}%
  \BibitemOpen
  \bibfield  {author} {\bibinfo {author} {\bibfnamefont {Y.-Y.}\ \bibnamefont
  {Chen}}, \bibinfo {author} {\bibfnamefont {P.}~\bibnamefont {Zhang}},
  \bibinfo {author} {\bibfnamefont {W.}~\bibnamefont {Zheng}}, \bibinfo
  {author} {\bibfnamefont {Z.}~\bibnamefont {Wu}}, \ and\ \bibinfo {author}
  {\bibfnamefont {H.}~\bibnamefont {Zhai}},\ }\href
  {https://arxiv.org/abs/1909.05183} {\bibinfo {title} {\emph {Many-Body
  Echo}}},\ \bibinfo {note} {arXiv:1909.05183}\BibitemShut {NoStop}%
\bibitem [{\citenamefont {Lewis~Jr}\ and\ \citenamefont
  {Riesenfeld}(1969)}]{lewis1969exact}%
  \BibitemOpen
  \bibfield  {author} {\bibinfo {author} {\bibfnamefont {H.~R.}\ \bibnamefont
  {Lewis~Jr}}\ and\ \bibinfo {author} {\bibfnamefont {W.}~\bibnamefont
  {Riesenfeld}},\ }\bibfield  {title} {\bibinfo {title} {\emph {An exact
  quantum theory of the time-dependent harmonic oscillator and of a charged
  particle in a time-dependent electromagnetic field}},\ }\href@noop {}
  {\bibfield  {journal} {\bibinfo  {journal} {Journal of Mathematical Physics}\
  }\textbf {\bibinfo {volume} {10}},\ \bibinfo {pages} {1458} (\bibinfo {year}
  {1969})}\BibitemShut {NoStop}%
\bibitem [{\citenamefont {Lewis~Jr}(1967)}]{lewis1967classical}%
  \BibitemOpen
  \bibfield  {author} {\bibinfo {author} {\bibfnamefont {H.~R.}\ \bibnamefont
  {Lewis~Jr}},\ }\bibfield  {title} {\bibinfo {title} {\emph {Classical and
  quantum systems with time-dependent harmonic-oscillator-type Hamiltonians}},\
  }\href@noop {} {\bibfield  {journal} {\bibinfo  {journal} {Physical Review
  Letters}\ }\textbf {\bibinfo {volume} {18}},\ \bibinfo {pages} {510}
  (\bibinfo {year} {1967})}\BibitemShut {NoStop}%
\bibitem [{mon()}]{monopole_line}%
  \BibitemOpen
  \href@noop {} {}\bibinfo {note} {The parametrization of $\mathbf{u}$ can be
  viewed as a U(1) fibration of SU(1,1). The non-compactness of SU(1,1)
  naturally leads to a non-compact base space (the ``Bloch'' hyperboloid shown
  in Fig. 1). As a consequence, the corresponding Berry curvature remains
  finite on the base space, which means the monopole line has to be infinitely
  long.}\BibitemShut {Stop}%
\bibitem [{\citenamefont {Nguyen}\ \emph {et~al.}(2019)\citenamefont {Nguyen},
  \citenamefont {Tsatsos}, \citenamefont {Luo}, \citenamefont {Lode},
  \citenamefont {Telles}, \citenamefont {Bagnato},\ and\ \citenamefont
  {Hulet}}]{nguyen2019parametric}%
  \BibitemOpen
  \bibfield  {author} {\bibinfo {author} {\bibfnamefont {J.}~\bibnamefont
  {Nguyen}}, \bibinfo {author} {\bibfnamefont {M.}~\bibnamefont {Tsatsos}},
  \bibinfo {author} {\bibfnamefont {D.}~\bibnamefont {Luo}}, \bibinfo {author}
  {\bibfnamefont {A.}~\bibnamefont {Lode}}, \bibinfo {author} {\bibfnamefont
  {G.~D.}\ \bibnamefont {Telles}}, \bibinfo {author} {\bibfnamefont {V.~S.}\
  \bibnamefont {Bagnato}}, \ and\ \bibinfo {author} {\bibfnamefont
  {R.}~\bibnamefont {Hulet}},\ }\bibfield  {title} {\bibinfo {title} {\emph
  {Parametric excitation of a Bose-Einstein condensate: from Faraday waves to
  granulation}},\ }\href@noop {} {\bibfield  {journal} {\bibinfo  {journal}
  {Physical Review X}\ }\textbf {\bibinfo {volume} {9}},\ \bibinfo {pages}
  {011052} (\bibinfo {year} {2019})}\BibitemShut {NoStop}%
\bibitem [{der()}]{derivation_3rd_eq}%
  \BibitemOpen
  \href@noop {} {}\bibinfo {note} {Label the three equations in
  Eq.~\eqref{linear_eq} by \textcircled{1}, \textcircled{2} and
  \textcircled{3}, they can be reduced to a single equation by considering
  $\frac{d^2}{dt^2}(\text{\textcircled{1}}-\text{\textcircled{3}})-2\epsilon_\mathbf{k}
  \frac{d}{dt}\text{\textcircled{2}}+4\epsilon_\mathbf{k}
  gn\text{\textcircled{1}}-4\epsilon_\mathbf{k}(\epsilon_\mathbf{k}+gn)\text{\textcircled{3}}$.}\BibitemShut
  {Stop}%
\bibitem [{\citenamefont {Deng}\ \emph {et~al.}(2016)\citenamefont {Deng},
  \citenamefont {Shi}, \citenamefont {Diao}, \citenamefont {Yu}, \citenamefont
  {Zhai}, \citenamefont {Qi},\ and\ \citenamefont {Wu}}]{deng2016observation}%
  \BibitemOpen
  \bibfield  {author} {\bibinfo {author} {\bibfnamefont {S.}~\bibnamefont
  {Deng}}, \bibinfo {author} {\bibfnamefont {Z.-Y.}\ \bibnamefont {Shi}},
  \bibinfo {author} {\bibfnamefont {P.}~\bibnamefont {Diao}}, \bibinfo {author}
  {\bibfnamefont {Q.}~\bibnamefont {Yu}}, \bibinfo {author} {\bibfnamefont
  {H.}~\bibnamefont {Zhai}}, \bibinfo {author} {\bibfnamefont {R.}~\bibnamefont
  {Qi}}, \ and\ \bibinfo {author} {\bibfnamefont {H.}~\bibnamefont {Wu}},\
  }\bibfield  {title} {\bibinfo {title} {\emph {Observation of the Efimovian
  expansion in scale-invariant Fermi gases}},\ }\href@noop {} {\bibfield
  {journal} {\bibinfo  {journal} {Science}\ }\textbf {\bibinfo {volume}
  {353}},\ \bibinfo {pages} {371} (\bibinfo {year} {2016})}\BibitemShut
  {NoStop}%
\bibitem [{\citenamefont {Shi}\ \emph {et~al.}(2017)\citenamefont {Shi},
  \citenamefont {Qi}, \citenamefont {Zhai},\ and\ \citenamefont
  {Yu}}]{shi2017dynamic}%
  \BibitemOpen
  \bibfield  {author} {\bibinfo {author} {\bibfnamefont {Z.-Y.}\ \bibnamefont
  {Shi}}, \bibinfo {author} {\bibfnamefont {R.}~\bibnamefont {Qi}}, \bibinfo
  {author} {\bibfnamefont {H.}~\bibnamefont {Zhai}}, \ and\ \bibinfo {author}
  {\bibfnamefont {Z.}~\bibnamefont {Yu}},\ }\bibfield  {title} {\bibinfo
  {title} {\emph {Dynamic super Efimov effect}},\ }\href@noop {} {\bibfield
  {journal} {\bibinfo  {journal} {Physical Review A}\ }\textbf {\bibinfo
  {volume} {96}},\ \bibinfo {pages} {050702} (\bibinfo {year}
  {2017})}\BibitemShut {NoStop}%
\bibitem [{Lya()}]{Lyapunov}%
  \BibitemOpen
  \href@noop {} {}\bibinfo {note} {Note that the Lyapunov exponent for
  Eq.~\eqref{2nd_eq} is half the exponent for Eq.~\eqref{3rd_eq}.}\BibitemShut
  {Stop}%
\bibitem [{\citenamefont {McLachlan}(1951)}]{mclachlan1951theory}%
  \BibitemOpen
  \bibfield  {author} {\bibinfo {author} {\bibfnamefont {N.~W.}\ \bibnamefont
  {McLachlan}},\ }\bibfield  {title} {\bibinfo {title} {\emph {Theory and
  application of Mathieu functions}},\ }\href@noop {} {\  (\bibinfo {year}
  {1951})}\BibitemShut {NoStop}%
\bibitem [{\citenamefont {Conduit}(2012)}]{conduit2012line}%
  \BibitemOpen
  \bibfield  {author} {\bibinfo {author} {\bibfnamefont {G.}~\bibnamefont
  {Conduit}},\ }\bibfield  {title} {\bibinfo {title} {\emph {Line of Dirac
  monopoles embedded in a Bose-Einstein condensate}},\ }\href@noop {}
  {\bibfield  {journal} {\bibinfo  {journal} {Physical Review A}\ }\textbf
  {\bibinfo {volume} {86}},\ \bibinfo {pages} {021605} (\bibinfo {year}
  {2012})}\BibitemShut {NoStop}%
\end{thebibliography}%
\end{document}